\documentclass[12pt]{iopart}
\usepackage{iopams}
\usepackage{setstack}
\usepackage{graphicx}
\usepackage{CJK}
\begin{document}
	
\title[]{Understanding current-driven dynamics of magnetic N\'{e}el walls in heavy metal/ferromagnetic metal/oxide trilayers}

\author{Mei Li$^{1,2}$, Jianbo Wang$^{2,4}$ and Jie Lu$^{3,4}$}

\address{$^1$ Physics Department, Shijiazhuang University, Shijiazhuang, Hebei 050035, China}
\address{$^2$ School of Physics and Technology, Center for Electron Microscopy and MOE Key Laboratory of Artificial Micro- and Nano-structures, Wuhan University, Wuhan 430072, China}
\address{$^3$ College of Physics and Information Engineering, Hebei Advanced Thin Films Laboratory, Hebei Normal University, Shijiazhuang 050024, China}
\address{$^4$ Authors to whom any correspondence should be addressed.}

\ead{wang@whu.edu.cn,jlu@hebtu.edu.cn}


\begin{abstract}
We consider analytically current-driven dynamics of magnetic N\'{e}el walls in 
heavy metal/ferromagnetic metal/oxide trilayers where strong spin-orbit coupling
and interfacial Dzyaloshinskii-Moriya interaction (i-DMI) coexist.
We show that field-like spin-orbit torque (FL-SOT)
with effective field along $\mathbf{n}\times\hat{\mathbf{J}}$ 
($\mathbf{n}$ being the interface normal and $\hat{\mathbf{J}}$ being the charge current direction)
and i-DMI induced torque can both lead to Walker breakdown suppression meanwhile leaving the 
wall mobility (velocity versus current density) unchanged. 
However, i-DMI itself can not induce the ``universal absence of Walker breakdown" (UAWB) 
while FL-SOT exceeding a certain threshold can. 
Finitely-enlarged Walker limits before UAWB are theoretically calculated and well explain existing data.
In addition, change in wall mobility and even its sign-inversion
can be understood only if the anti-damping-like (ADL) SOT is appended.
For N\'{e}el walls in ferromagnetic-metal layer with both perpendicular and in-plane anisotropies,
we have calculated the respective modifications of wall mobility 
under the coexistence of spin-transfer torque, SOTs and i-DMI.
Analytics shows that in trilayers with perpendicular anisotropy strong enough spin Hall angle
and appropriate sign of i-DMI parameter can lead to sign-inversion in wall mobility even under small enough current density,
while in those with in-plane anisotropy this only occurs for current density in a specific range.
\end{abstract}


\submitto{\NJP}

\maketitle


\section{Introduction}\label{section:introduction}
Pure current-induced domain wall propagation in magnetic nanostructures has attracted intensive
attention for decades starting from academic interests in understanding the interplay between itinerant
spinful electrons and localized magnetic
moments\cite{Berger_PRB_1996,Slonczewski_JMMM_1996,SCZhang_PRB_1998,Zangwill_PRB_2002}.
In monolayer ferromagnetic nanostrips,
in-plane currents drive domain walls to propagate along the direction of electron flow
through the spin transfer process\cite{Tatara_PRL_2004,SZhang_PRL_2004a,SZhang_PRL_2004b,Thiaville_EPL_2005,PYan_EPL_2010,ZZSun_EPJB_2011,
	APL_83_509_2003,PRL_92_077205_2004,PRL_95_026601_2005,AdvPhys_54_585_2005,PRL_98_037204_2007},
which leads to promising applications in future
magnetic racetrack memories\cite{Science_320_190_2008,NatNanotech_10_195_2015},
shift registers\cite{Science_320_209_2008,NatNanotech_7_499_2012}
and memristors\cite{IEEE_30_294_2009,JAP_111_07D303_2012}, etc.
However, in these monolayers the wall velocity is at most $10^2\ \mathrm{m/s}$
even when the current density is up to $10^{8}\ \mathrm{A/cm^2}$.
This comes from the fact that spin transfer torques (STTs) therein
can not be strong as the exchange energy avoids abrupt changes in magnetization texture.
To improve the current efficiency, the current-perpendicular-to-plane (CPP)
configuration in narrow and long spin valves is proposed
\cite{Cros_PRL_2009,Boone_PRL_2010,Cros_NatPhys_2011}:
to reach the same velocity level ($10^2\ \mathrm{m/s}$), the current density
for ``planar polarizer" case is reduced to $10^{7}\ \mathrm{A/cm^2}$ while that for
``perpendicular polarizer" can even be lowered to $10^{6}\ \mathrm{A/cm^2}$.
However, the rapidly increasing CPP cross-section area largely offsets the decrease in current density.
Even if the current is forced to focus on wall region, precise dynamical synchronization in
real experiments remains challenging.

Over the past decade, in heavy metal/ferromagnetic metal/oxide (HM/FMM/Oxide) trilayers,
axial domain wall propagation in FMM layer with perpendicular magnetic anisotropy (PMA) 
or in-plane magnetic anisotropy (IPMA) driven by axial currents are
experimentally observed\cite{Miron_Nmat_2010,Seo_APL_2010,Buhrman_Science_2012,Buhrman_PRL_2012,Garello_NatNanoTech_2013,
	Buhrman_PRL_2016,Buhrman_PRB_2016,Ghosh_PRAppl_2017,Moore_APL_2008,Miron_Nature_2011,Beach_Nmat_2013,
	Lee_PRL_2011,Miron_Nmat_2011,Beach_APL_2012,Stuart_APE_2012,Koopmans_APL_2012,Parkin_NatNanoTech_2013}.
In certain case ($\mathrm{Pt/Co/AlO_x}$), walls can move at a high velocity up to $400\ \mathrm{m/s}$
when current density is around $10^{8}\ \mathrm{A/cm^2}$\cite{Miron_Nmat_2011}.
More interestingly, walls with certain polarity can even move in the
direction of charge current\cite{Moore_APL_2008,Miron_Nature_2011,Beach_Nmat_2013,Lee_PRL_2011,Miron_Nmat_2011,
	Beach_APL_2012,Stuart_APE_2012,Koopmans_APL_2012,Parkin_NatNanoTech_2013},
which is also confirmed by numerical simulations
\cite{Linder_PRB_2013a,Martinez_AIPAdv_2013,Miron_JAP_2014,Stier_PRB_2014}.
To understand these findings, spin-orbit torques (SOTs) from strong
spin-orbit coupling (SOC) in these trilayers are proposed
\cite{Zhang_PRB_2008,Zhang_PRB_2009,Matos_PRB_2009,Miron_RoyalSocietyA_2011,Manchon_PRB_2013a,Manchon_PRB_2013b,
	Manchon_PRL_2012,Seo_PRB_2012,Pesin_PRB_2012,Bijl_PRB_2012,Qaiumzadeh_PRB_2015,Kurebayashi_NatNanoTech_2014,Li_PRB_2015,
	ChenWei_PRL_2015}.
Suppose $\hat{\mathbf{J}}$ is the charge current direction and
$\mathbf{n}$ is interface normal.
Mathematically, SOTs can be decomposed into two perpendicular components:
(a) $\propto\mathbf{m}\times(\hat{\mathbf{J}}\times\mathbf{n})$ which is odd
in magnetization direction $\mathbf{m}$ and usually referred to as field-like (FL) torque;
(b) $\propto\mathbf{m}\times[\mathbf{m}\times(\hat{\mathbf{J}}\times\mathbf{n})]$
which is even in $\mathbf{m}$ and usually called anti-damping-like (ADL) torque.
Physically, two typical mechanisms are of most importance:
the ``spin Hall torques" from the spin Hall effect (SHE)\cite{Hirsch_PRL_1999} in HM layer and
the ``Rashba torques" from the structure inversion asymmetry (SIA) at the HM/FMM interface.
In early literatures, ADL-SOTs are believed to stem mostly from bulk SHE
while FL-SOTs are mainly attributed to interfacial Rashba SOC.
However, recent works based on scattering-related mechanisms\cite{Manchon_PRL_2012,Seo_PRB_2012,Pesin_PRB_2012,Bijl_PRB_2012,Qaiumzadeh_PRB_2015}
and intrinsic Berry curvature\cite{Kurebayashi_NatNanoTech_2014,Li_PRB_2015}
reveal that Rashba SOC can cause FL and ADL SOTs with similar strength.
Meantime, quantum tunneling of spin current from HMs to FMMs\cite{ChenWei_PRL_2015} allows
SHE to provide FL and ADL SOTs with comparable magnitude.
So far, physical source of SOTs is still a hot issue under debate\cite{Freimuth_PRB_2014,Stiles_PRB__2016a,Stiles_PRB__2016b,Ado_PRB_2017}.
Besides Rashba SOT, SIA in these trilayers also leads to the 
interfacial Dzyaloshinskii-Moriya interaction (i-DMI)\cite{Dzyaloshinskii_1957,Moriya_1957,Dzyaloshinskii_1965}, 
which favors a canting of the spins and stabilizes 
N\'{e}el walls in FMM layers.

Meantime, analytics with
Lagrangian functional\cite{Linder_PRB_2013b,Tatara_PRB_2008,He_JAP_2013}
and simulations\cite{Martinez_JAP_2012,Martinez_ACMP_2012,Seo_JMMM_2012,Seo_APL_2012,Cros_PRB_2013,Martinez_APL_2013,Martinez_JAP_2014}
based on Landau-Lifshitz-Gilbert (LLG) dynamical equation\cite{LLG_equation}
have been performed to explain N\'{e}el wall dynamics in HM/FMM/Oxide trilayers
in the framework of one-dimensional collective coordinate model (1D-CCM).
All these works focus on two novel features in experiments:
(i) Walker breakdown suppression thus high wall velocity and (ii) wall motion opposed to electron flow
and the corresponding ``polarity sensitivity".
Historically, the Rashba-SOC-induced FL-SOT is first proposed to explain both novelties\cite{Martinez_JAP_2012,Martinez_ACMP_2012,Seo_JMMM_2012}.
In addition, the novelty (ii) is also reproduced numerically by only ADL-SOTs from SHE in IPMA systems\cite{Seo_APL_2012}.
In 2017, Risingg{\aa}d and Linder proposed the ``universal absence of Walker breakdown" (UAWB)
of N\'{e}el walls in PMA systems for strong enough
Rashba or SHE effect\cite{Risinggad_PRB_2017} with the coexistence of i-DMI.
However, to our knowledge there are no explicit analytical expressions for finitely enlarged Walker limit
before the occurrence of UAWB.
In addition, theoretical criteria for sign-inversion in wall mobility (velocity versus current density) 
and the corresponding ``polarity selection rule"
for trilayers with both PMA and IPMA are absent.
The role of i-DMI in all these processes is also unclear.
Explorations to these issues constitute the main content of this paper.

The rest of this paper is organized as follows.
In \sref{Modeling} the system set up and its modelization are briefly introduced. 
Also, the static N\'{e}el wall configurations and the dynamical equations for 
FMM layers with both PMA and IPMA are presented.
Then in \sref{section:WalkerBreakdownSuppression} within 1D-CCM we provide analytical expressions
of finitely enlarged Walker limits by FL-SOT and/or i-DMI induced torque before UAWB.
In \sref{MobilityChangeByADLSOT}, theoretical criteria for sign-inversion in wall mobility and 
the corresponding ``polarity selection rule"
for FMM layers with both PMA and IPMA are provided under the coexistence of ADL-SOT and i-DMI.
Finally, concluding remarks are provided in the last section.

\section{Modeling and preparations}\label{Modeling}
We consider an HM/FMM/Oxide trilayer with a domain wall formed in FMM layer.
Generally, the FMM layer has strong PMA or IPMA.
Typical example for the former (latter) case is Co (NiFe).  
Meanwhile, the HM layer is composed of Pt or Ta in most experiments.
For both cases, the in-plane charge current flows along the long axis of the strip
with density $j_a$.
As passing through the trilayer, the charge current splits into two parts.
Suppose $j_{\mathrm{F}}$ ($j_{\mathrm{H}}$) to be the current density in FMM (HM) layer.
A simple circuit model tells us that $j_{\mathrm{F}}=j_a(t_{\mathrm{F}}+t_{\mathrm{H}})\sigma_{\mathrm{F}}/(t_{\mathrm{F}}\sigma_{\mathrm{F}}+t_{\mathrm{H}}\sigma_{\mathrm{H}})$ and $j_{\mathrm{H}}=j_a(t_{\mathrm{F}}+t_{\mathrm{H}})\sigma_{\mathrm{H}}/(t_{\mathrm{F}}\sigma_{\mathrm{F}}+t_{\mathrm{H}}\sigma_{\mathrm{H}})$, where $t_{\mathrm{F}}$ ($t_{\mathrm{H}}$) and $\sigma_{\mathrm{F}}$ ($\sigma_{\mathrm{H}}$) 
are the thickness and conductivity of the FMM (HM) layer, respectively.
For the most common FMM (Co, Ni, Fe) and HM (Pt, Ta, Ir) materials, 
the conductivity varies from 10 to 20 $\mathrm{(\mu\Omega m)}^{-1}$.
For simplicity in this work, we set $\sigma_{\mathrm{F}}=\sigma_{\mathrm{H}}$ thus 
$j_{\mathrm{F}}=j_{\mathrm{H}}=j_a$.

\begin{figure}[htbp]
	\centering
	\includegraphics[width=0.7\textwidth]{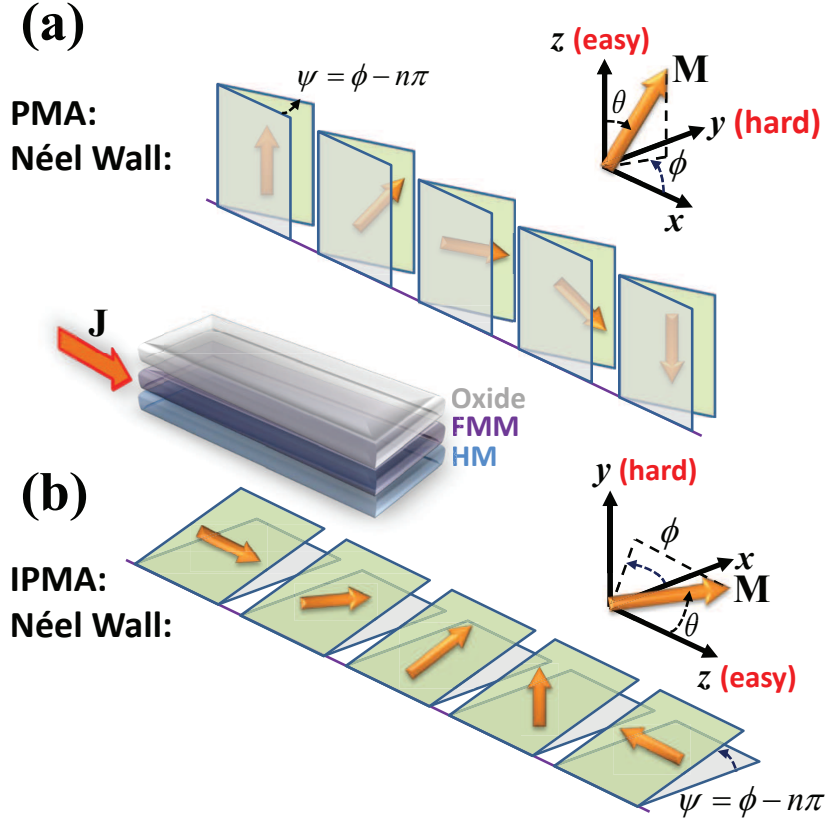}
	\caption{(Color online) Sketch of a typical HM/FMM/Oxide trilayer in which
		a N\'{e}el wall is formed in FMM layer with (a) PMA and (b) IPMA, as a result of
		energy minimization.
		The corresponding coordinate system is depicted at the up-right corner in each subfigure.	
		In each case, gray planes describe the planar $\phi-$distribution of static magnetization texture.
		When in-plane charge current $\mathbf{J}$ is applied, magnetization
		vectors will be driven to tilt from their static locations by $\psi$,
		as indicated by the green planes.
		This sketch is inspired by figure 1 in Ref. \cite{Linder_PRB_2013b}.}\label{fig1}
\end{figure}

For trilayers with PMA, the coordinate system is depicted in \fref{fig1}a:
$\mathbf{e}_x$ is along the long axis of strip in which charge current flows, 
$\mathbf{e}_z$ is the interface normal
and $\mathbf{e}_y=\mathbf{e}_z\times\mathbf{e}_x$.
The easy (hard) axis of the FMM layer with PMA lies in $\mathbf{e}_z$ ($\mathbf{e}_y$) direction.
While for trilayers with IPMA (see \fref{fig1}b), $\mathbf{e}_z(\mathbf{e}_y)$ is along the long axis (interface normal) 
of the strip which is the easy (hard) axis, and $\mathbf{e}_x=\mathbf{e}_y\times\mathbf{e}_z$.

\subsection{Dynamical equation}
In thin enough strips, most of the nonlocal magnetostatic energy can be described by local
quadratic terms of $M_{x,y,z}$ via three average demagnetization factors.
Thus in the absence of any external magnetic field, the total magnetic energy density functional 
takes the following form
\begin{equation}\label{E_tot_density}
\mathcal{E}_{\mathrm{tot}}[\mathbf{M}] =\mathcal{E}_{\mathrm{ex}}+\mathcal{E}_{\mathrm{DM}}+\mathcal{E}_{\mathrm{ani}},
\end{equation}
in which $\mathcal{E}_{\mathrm{ex}}=J(\nabla\mathbf{m})^2$ and $J(>0)$ is the exchange stiffness. 
The i-DMI contribution is $\mathcal{E}_{\mathrm{DM}}=D[(\mathbf{m}\cdot\mathbf{n})(\nabla\cdot \mathbf{m})-(\mathbf{m}\cdot\nabla)(\mathbf{m}\cdot\mathbf{n})]$, where $D$ is the magnitude of
i-DMI vector and $\mathbf{n}$ is the interface normal. 
The total magnetic anisotropy energy density is 
$\mathcal{E}_{\mathrm{ani}}=(\mu_0 M_s^2/2)[-k_{\mathrm{E}}(\mathbf{m}\cdot\mathbf{n}_{\mathrm{E}})^2 +k_{\mathrm{H}} (\mathbf{m}\cdot\mathbf{n}_{\mathrm{H}})^2]$, 
in which $\mathbf{n}_{\mathrm{E}}(\mathbf{n}_{\mathrm{H}})$ and $k_{\mathrm{E}}(k_{\mathrm{H}})$ are the unit vector
and total anisotropy coefficient in easy (hard) axis, respectively.
The time evolution of magnetization texture
$\mathbf{M}(\mathbf{r},t)\equiv M_s\mathbf{m}(\mathbf{r},t)$
with fixed saturation magnetization $M_s$ is governed by the generalized
LLG equation
\begin{equation}\label{LLG_vector}
\frac{\partial \mathbf{m}}{\partial t}=-\gamma \mathbf{m}\times \mathbf{H}_{\mathrm{eff}}
+\alpha\left(\mathbf{m}\times\frac{\partial \mathbf{m}}{\partial t}\right)
+\mathbf{T}_{\mathrm{STT}}+\mathbf{T}_{\mathrm{SOT}},
\end{equation}
where $\mathbf{H}_{\mathrm{eff}}=-(\delta \mathcal{E}_{\mathrm{tot}}/\delta \mathbf{m})/(\mu_{0}M_s)$
is the effective field, $\gamma$ and $\alpha$ are the gyromagnetic ratio and
phenomenological damping coefficient, respectively.

Note $\mathbf{T}_{\mathrm{STT}}$ only appears for inhomogeneous magnetization texture with\cite{SZhang_PRL_2004a,SZhang_PRL_2004b}
\begin{equation}\label{SST_trilayer}
\mathbf{T}_{\mathrm{STT}}=B_J\frac{\partial\mathbf{m}}{\partial \hat{\mathbf{J}}}-\beta B_J\mathbf{m}\times\frac{\partial\mathbf{m}}{\partial \hat{\mathbf{J}}},
\end{equation}
where $B_J=g_e\mu_B P j_{\mathrm{F}} /(2e M_s)\approx \mu_B P j_a/(e M_s)$, with $e,g_e,\mu_B$ being the absolute value of electron charge,
the electron $g-$factor and Bohr magneton, respectively.
$P$ is the spin polarization of $j_{\mathrm{F}}$.
The two terms in the right hand side of equation \eref{SST_trilayer} are the so-called adiabatic 
and non-adiabatic STTs, respectively.
They are the continuous counterparts of the Slonczeswki\cite{Slonczewski_JMMM_1996} and FL STTs in spin valves.
$\beta$ is the dimensionless coefficient
describing the relative strength of the nonadiabatic STT and usually of the same order as $\alpha$.
Previous works have verified that in traveling-wave mode STT-driven walls always move in the direction
of electron flow, which is attributed to the existence of nonadiabatic ingredient ($\beta-$term).

Generally SOTs have both FL and ADL components.
Each component includes the contributions from both SHE and Rashba SOC.
In this work we focus on domain wall dynamics rather than physical sources of SOTs,
thus $\mathbf{T}_{\mathrm{SOT}}$ can be written as
\begin{equation}\label{SOT_trilayer}
\mathbf{T}_{\mathrm{SOT}}=-\gamma H_{\mathrm{FL}}\mathbf{m}\times\left(\mathbf{n}\times\hat{\mathbf{J}}\right)
-\gamma H_{\mathrm{ADL}}\mathbf{m}\times\left[\mathbf{m}\times\left(\mathbf{n}\times\hat{\mathbf{J}}\right)\right].
\end{equation}
Both $H_{\mathrm{FL}}$ and $H_{\mathrm{ADL}}$ stem from various physical processes
and have the unit of magnetic field.
Their ratio varies in a wide range for different trilayer systems.

\subsection{Static wall configuration}\label{Modeling:StaticWall}
In the absence of external charge current, the magnetization texture
eventually evolves into some equilibrium state. The ground state
is the one with a single domain which is of little interest.
Alternatively, the metastable state with a wall separating
two magnetic domains is of great importance for both academic and
industrial interests. In this subsection, we provide
static wall configurations for trilayers with both PMA and IPMA.

First we focus on FMM layers with PMA (see \fref{fig1}a), thus
$\mathbf{n}=\mathbf{n}_{\mathrm{E}}=\mathbf{e}_z$ and $\mathbf{n}_{\mathrm{H}}=\mathbf{e}_y$.
For statics, the magnetization is no longer function of time but only
varies with location along $x-$axis in 1D-CCM.
By dropping a constant $-k_{\mathrm{E}}\mu_0 M_s^2/2$, the total energy density turns to
\begin{equation}\label{E_tot_density_static_1}
\eqalign{
 \mathcal{E}_{\mathrm{PMA}}[\mathbf{M}] = \frac{1}{2}\mu_0
 M_s^2\sin^2\theta\left(k_{\mathrm{E}}+k_{\mathrm{H}}\sin^2\phi\right)+J\left[\left(\theta'\right)^2+\sin^2\theta\left(\phi'\right)^2\right] \cr
 \qquad \qquad \qquad +D\left(\cos\phi\theta'-\sin\theta\cos\theta\sin\phi\phi'\right),
}
\end{equation}
where $\theta$ ($\phi$) is the polar (azimuthal) angle of the magnetization vector (see \fref{fig1}a)
and a ``prime" means $\mathrm{d}/\mathrm{d}x$.
Physically, a static wall configuration should provide a minimum of the total magnetic energy.
For this purpose, first we should have $\phi'\equiv 0$ to suppress the exchange energy,
which makes $\phi$ a collective coordinate.
Then we introduce the Lagrangian functional $L=\int\mathcal{L}\mathrm{d}^3 \mathbf{r}$ with Lagrangian density
\begin{equation}\label{Lagrangian_density}
\mathcal{L}=\frac{\mu_0 M_s}{\gamma}\frac{\partial\phi}{\partial t}(1-\cos\theta)-\mathcal{E}_{\mathrm{tot}}[\mathbf{M}],
\end{equation}
and the boundary condition
\begin{equation}\label{theta_static_BC}
\mathbf{m}(x=\mp\infty)=\pm\eta\mathbf{e}_z,
\end{equation}
with $\eta=\pm 1$ coming from the two-fold symmetry of magnetic anisotropy in easy axis
and can be viewed as the topological charge of this wall.
The corresponding Euler equation
\begin{equation}\label{Euler_equation_general}
\frac{\mathrm{d}}{\mathrm{d}x}\left(\frac{\partial \mathcal{L}}{\partial (\mathrm{d}\theta/\mathrm{d}x)}\right)-\frac{\mathrm{d}}{\mathrm{d}\theta}\mathcal{L}=0
\end{equation}
together with the static condition $\partial\phi/\partial t=0$ lead to
\begin{equation}\label{Euler_equation_static}
\frac{\mathrm{d}^2\theta}{\mathrm{d}x^2}=\frac{\sin\theta\cos\theta}{\Delta^2(\phi)},\quad \Delta(\phi)=\sqrt{\frac{2J}{\mu_0 M_s^2\left(k_{\mathrm{E}}+k_{\mathrm{H}}\sin^2\phi\right)}}.
\end{equation}
Its soliton solution is the well-known Walker profile\cite{Walker}
\begin{equation}\label{Walker_profile_static}
\ln\tan\frac{\theta}{2}=\eta\frac{x-x_0}{\Delta(\phi)},
\end{equation}
where $x_0$ denotes the wall center position.
Putting it back into \eref{E_tot_density_static_1}, one has
\begin{equation}\label{E_tot_density_static_2}
	\mathcal{E}_{\mathrm{PMA}}[\mathbf{M}] = \mu_0
	M_s^2\sin^2\theta\left(k_{\mathrm{E}}+k_{\mathrm{H}}\sin^2\phi\right)+\eta D\cos\phi\frac{\sin\theta}{\Delta(\phi)}.
\end{equation}
The energy minimization strategy then naturally demands that $\sin\phi= 0$ and $\cos\phi=-\eta \mathrm{sgn}(D)$, 
where ``sgn" is the sign function. This means that in PMA systems, the i-DMI favors N\'{e}el wall and 
further selects wall polarity (sign of $\left\langle m_y\right\rangle$).

Next we turn to IPMA systems (see \fref{fig1}b) in which $\mathbf{n}=\mathbf{n}_{\mathrm{H}}=\mathbf{e}_y$ and $\mathbf{n}_{\mathrm{E}}=\mathbf{e}_z$.
After similar process, we obtain the same $\theta-$profile as in equation \eref{Walker_profile_static} 
except for $x (x_0)\rightarrow z (z_0)$, 
under which the total energy density of IPMA systems becomes
\begin{equation}\label{E_tot_density_static_IPMA}
\mathcal{E}_{\mathrm{IPMA}}[\mathbf{M}] = \mu_0
M_s^2\sin^2\theta\left(k_{\mathrm{E}}+k_{\mathrm{H}}\sin^2\phi\right)-\eta D\sin\phi\frac{\sin\theta}{\Delta(\phi)}.
\end{equation}
To minimize the first term in the right hand side of the above equation, one also need $\sin\phi(z)\equiv 0$ which 
eliminates the i-DMI term. This implies that in IPMA systems, the N\'{e}el wall is naturally the
result of energy minimization strategy and the i-DMI does not select wall polarity.

\subsection{General scalar LLG equations}
By taking into account the conversion between Descartes and spherical coordinate systems,
the vectorial LLG equation \eref{LLG_vector} is transformed into the following scalar pair
\begin{equation}\label{LLG_scalar_general_1}
	\eqalign{
		\left(\dot{\theta}+\alpha\sin\theta\dot{\phi}\right)-B_J \left(\theta'+\beta\sin\theta\phi'\right)=\gamma \tilde{A}, \cr
		\left(\sin\theta\dot{\phi}-\alpha\dot{\theta}\right)-B_J \left(\sin\theta\phi'-\beta\theta'\right)=\gamma \tilde{B}.
	 }
\end{equation}

For PMA systems, one has
\begin{equation}\label{LLG_scalar_general_2}
	\eqalign{
		\tilde{A}=A + H_{\mathrm{FL}}\cos\phi + H_{\mathrm{ADL}}\cos\theta\sin\phi, \cr
		\tilde{B}=B + H_{\mathrm{ADL}}\cos\phi - H_{\mathrm{FL}}\cos\theta\sin\phi,
	}
\end{equation}
with
\begin{equation}\label{AB_exact}
	\eqalign{
		A =  \frac{2D}{\mu_0 M_s}\sin\theta\sin\phi\theta'+\frac{2J}{\mu_0 M_s\sin\theta}(\phi'\sin^2\theta)'-k_{\mathrm{H}} M_s \sin\theta\sin\phi\cos\phi,   \cr
		B =  M_s\sin\theta\cos\theta(k_{\mathrm{E}}+k_{\mathrm{H}}\sin^2\phi) +\frac{2D}{\mu_0 M_s}\sin^2\theta\sin\phi\phi' \cr
		\qquad \qquad -\frac{2J}{\mu_0 M_s}\left[ \theta''-(\phi')^2\sin\theta\cos\theta \right],
	}
\end{equation}
where a ``dot (prime)" means $\partial/\partial t$ ($\partial/\partial x$).

While for IMPA system, alternatively one has
\begin{equation}\label{LLG_scalar_general_IPMA}
\eqalign{
	\tilde{A}=A - H_{\mathrm{FL}}\sin\phi + H_{\mathrm{ADL}}\cos\theta\cos\phi, \cr
	\tilde{B}=B - H_{\mathrm{ADL}}\sin\phi - H_{\mathrm{FL}}\cos\theta\cos\phi,
}
\end{equation}
with
\begin{equation}\label{AB_exact_IPMA}
\eqalign{
	A =  \frac{2D}{\mu_0 M_s}\sin\theta\cos\phi\theta'+\frac{2J}{\mu_0 M_s\sin\theta}(\phi'\sin^2\theta)'-k_{\mathrm{H}} M_s \sin\theta\sin\phi\cos\phi,   \cr
	B =  M_s\sin\theta\cos\theta(k_{\mathrm{E}}+k_{\mathrm{H}}\sin^2\phi) +\frac{2D}{\mu_0 M_s}\sin^2\theta\cos\phi\phi' \cr
	\qquad \qquad -\frac{2J}{\mu_0 M_s}\left[ \theta''-(\phi')^2\sin\theta\cos\theta \right],
}
\end{equation}
in which a ``prime" means $\partial/\partial z$.

\section{Walker breakdown suppression by FL-SOT and/or i-DMI}\label{section:WalkerBreakdownSuppression}
In this section we present analytical expressions of finitely enlarged Walker limit before UAWB
in the absence of ADL-SOT.
We will show the different roles of FL-SOT and i-DMI in modulating the STT-initiated traveling-wave mode of domain wall.

\subsection{Brief review of wall dynamics under pure STT}
For a N\'{e}el wall in an isolated FMM and driven by pure STTs from axial currents 
($H_{\mathrm{FL}}=H_{\mathrm{ADL}}\equiv 0$, $D\equiv 0$ and $B_J\ne 0$),
the static wall profile can be generalized to\cite{SZhang_PRL_2004a,SZhang_PRL_2004b}
\begin{equation}\label{Walker_profile_traveling_wave}
\ln\tan\frac{\theta}{2}=\frac{\eta}{\Delta(\phi)}\left[r-\int_0^t v(\tau)\mathrm{d}\tau\right], \quad \phi=\phi(t),
\end{equation}
where $r=x(z)$ for PMA (IPMA) case, $\Delta(\phi)$ is the same as in equation \eref{Euler_equation_static} 
and $v(t)$ is the wall velocity.
Then for trilayers with both PMA and IPMA, we have
\begin{equation}\label{AB_tilde_traveling_wave}
\tilde{A}=-k_{\mathrm{H}}M_s\sin\theta\sin\phi\cos\phi,\quad \tilde{B}=0.
\end{equation}
Putting back into the scalar LLG equations, one has
\begin{equation}\label{v_phi_dot_Walker_traveling_wave}
  \eqalign{
		\frac{v(t)}{\Delta(\phi)}=\frac{\eta\gamma }{2(1+\alpha^2)}H_{\mathrm{K}}\sin 2\phi - \frac{1+\alpha\beta}{1+\alpha^2}\frac{B_J}{\Delta(\phi)}, \cr
		\dot{\phi}=-\frac{\alpha\gamma }{2(1+\alpha^2)}H_{\mathrm{K}}\sin 2\phi +\frac{\alpha-\beta}{1+\alpha^2}\frac{\eta B_J}{\Delta(\phi)},
	}
\end{equation}
with $H_{\mathrm{K}}\equiv k_{\mathrm{H}}M_s$. By eliminating the ``$H_{\mathrm{K}}\sin 2\phi$" term, $v(t)$
and $\dot{\phi}$ are directly related as
\begin{equation}\label{v_phi_dot_direct_Walker_traveling_wave}
v(t)=-\frac{\eta\Delta(\phi)}{\alpha}\dot{\phi}-\frac{\beta}{\alpha} B_J.
\end{equation}
By setting $\dot{\phi}=0$ in the above equation, the Walker limit
\begin{equation}\label{WB}
J_{\mathrm{W}}\equiv\frac{eM_s}{\mu_B P}\cdot\frac{\Delta\gamma H_{\mathrm{W}}}{|\alpha-\beta|},\quad H_{\mathrm{W}}\equiv \frac{\alpha H_{\mathrm{K}}}{2} 
\end{equation}
is obtained as the result of constraint $|\sin 2\phi|\le 1$. 
Here we neglect the breathing effect of dynamical wall width $\Delta(\phi)$.
Equations \eref{v_phi_dot_Walker_traveling_wave} and \eref{v_phi_dot_direct_Walker_traveling_wave} show that
when $|J_e|\le J_{\mathrm{W}}$ the wall propagates along electron-flow direction in a traveling-wave mode
with the velocity $-\beta B_J/\alpha$ and the tilting angle
\begin{equation}\label{Tilting_angle_Walker_profile}
\phi=\phi_0+\frac{1}{2}\arcsin\left[\eta\mathrm{sgn}(\alpha-\beta)\frac{j_a}{J_\mathrm{W}}\right],
\end{equation}
where $\phi_0=\arccos[-\eta\mathrm{sgn}(D)]$ for PMA case and $\phi_0=k\pi,k\in\mathbb{Z}$ for IPMA case.
Obviously when $\alpha=\beta$, $J_{\mathrm{W}}=+\infty$ implying the occurrence of UAWB.

\subsection{General framework under the coexistence of STT, FL-SOT and i-DMI}\label{WalkerBreakdownSuppression:GeneralFramework}
First we focus on trilayers with PMA. 
As illustrated, a traveling wave described by equation \eref{Walker_profile_traveling_wave}
can always be adopted to perform analytics\cite{Risinggad_PRB_2017}.
Under this wall profile,
$\tilde{A}$ and $\tilde{B}$ in equation \eref{LLG_scalar_general_2} becomes
\begin{equation}\label{AB_tilde_only_FL_SOTs}
\eqalign{
    \tilde{A}=\frac{2\eta D}{\mu_{0} M_s\Delta}\sin^2\theta\sin\phi-\frac{H_{\mathrm{K}}}{2}\sin\theta\sin 2\phi+H_{\mathrm{FL}}\cos\phi,\cr
    \tilde{B}=-H_{\mathrm{FL}}\cos\theta\sin\phi.
}
\end{equation}
Putting back to the scalar LLG equations, and then integrating
over the whole strip ($\int_{-\infty}^{+\infty}\sin\theta\mathrm{d}x/\Delta \equiv \int_{0}^{\pi}\mathrm{d}\theta$),
one has
\begin{equation}\label{v_phi_dot_only_FL_SOTs}
	\eqalign{
		\frac{v(t)}{\Delta(\phi)}=\frac{\eta\gamma \mathcal{H}(H_{\mathrm{K}},H_{\mathrm{FL}},D,\phi)}{2(1+\alpha^2)} - \frac{1+\alpha\beta}{1+\alpha^2}\frac{B_J}{\Delta(\phi)}, \cr
		\dot{\phi}=-\frac{\alpha\gamma \mathcal{H}(H_{\mathrm{K}},H_{\mathrm{FL}},D,\phi)}{2(1+\alpha^2)} +\frac{\alpha-\beta}{1+\alpha^2}\frac{\eta B_J}{\Delta(\phi)},
	}
\end{equation}
with the functional
\begin{equation}\label{H_functional_only_FL_SOTs}
\mathcal{H}(H_{\mathrm{K}},H_{\mathrm{FL}},D,\phi)\equiv H_{\mathrm{K}}\sin 2\phi-\frac{\eta\pi D}{\mu_{0} M_s \Delta}\sin\phi-\pi H_{\mathrm{FL}}\cos\phi.
\end{equation}
Obviously, equation \eref{v_phi_dot_only_FL_SOTs} shares the same structure with
equation \eref{v_phi_dot_Walker_traveling_wave}, except for the substitution
of ``$H_{\mathrm{K}}\sin 2\phi$" by the functional $\mathcal{H}(H_{\mathrm{K}},H_{\mathrm{FL}},D,\phi)$,
thus leads to the rediscovery of equation \eref{v_phi_dot_direct_Walker_traveling_wave}.
\emph{This means neither the FL-SOT nor the i-DMI can change the wall mobility.
However they do suppress the Walker breakdown thus increase the upper limit of wall velocity 
in traveling-wave mode.}

To see this, we first define
\begin{equation}\label{a_b_definition}
 a\equiv \pi\frac{H_{\mathrm{FL}}}{H_{\mathrm{K}}}, \quad b\equiv \frac{\eta\pi D}{\mu_{0} M_s H_{\mathrm{K}} \Delta}.
\end{equation}
Then by setting $\dot{\phi}=0$, the second line in equation \eref{v_phi_dot_only_FL_SOTs} provides
\begin{equation}\label{phi_dot_0_only_FL_SOTs}
\sin 2\phi-a\cos\phi-b\sin\phi=\eta \mathrm{sgn}(\alpha-\beta)\frac{j_a}{J_\mathrm{W}}.
\end{equation}
Next we set $\left(H_{\mathrm{FL}}\right)_{\mathrm{W}}$ as the absolute effective field strength
when $|j_a|=J_{\mathrm{W}}$. Then by defining
\begin{equation}\label{a_W_only_FL_SOTs}
a_{\mathrm{W}}\equiv \frac{\pi \left(H_{\mathrm{FL}}\right)_{\mathrm{W}}}{ H_{\mathrm{K}}}>0,
\end{equation}
equation \eref{phi_dot_0_only_FL_SOTs} is rewritten as
\begin{equation}\label{Je_JW_only_FL_SOTs}
\frac{j_a}{J_{\mathrm{W}}} =\frac{\sin 2\phi-b\sin\phi}{\eta \mathrm{sgn}(\alpha-\beta)+a_{\mathrm{W}}\cos\phi}.
\end{equation}
By maximizing the absolute value of its right hand side, we get the modified Walker limit $J_{\mathrm{W}}^{\mathrm{FL+DM}}$.

If $a_{\mathrm{W}}>1$, $j_a/J_{\mathrm{W}}\rightarrow\infty$
when $\cos\phi=-\eta \mathrm{sgn}(\alpha-\beta)/a_{\mathrm{W}}$ 
thus leading to infinite $J_{\mathrm{W}}^{\mathrm{FL+DM}}$ (i.e. UAWB),
which is essentially the same as that from equation (8) of Ref. \cite{Risinggad_PRB_2017}.
When $a_{\mathrm{W}}=1$ and $b\ne 2$, 
without losing generality we set ``$\eta \mathrm{sgn}(\alpha-\beta)\equiv -1$".
As $\phi\rightarrow 0^+$ one has
$|j_a/J_{\mathrm{W}}|\approx |2\phi^{-1}(2-\phi^2-b)|\rightarrow +\infty$.
Note that a special parameter combination ``$a_{\mathrm{W}}=1$ and $b=2$"
leads to $|j_a/J_{\mathrm{W}}|=2|\sin\phi|$, which gives a doubled Walker limit.
Except for this special case, one would see that finite $J_{\mathrm{W}}^{\mathrm{FL+DM}}$ 
can only exist for $0<a_{\mathrm{W}}<1$.

On the other hand for trilayers with IPMA,
the only difference is that equation \eref{phi_dot_0_only_FL_SOTs} turns to
\begin{equation}\label{phi_dot_0_only_FL_SOTs_IPMA}
\sin 2\phi+a\sin\phi-b\cos\phi=\eta \mathrm{sgn}(\alpha-\beta)\frac{j_a}{J_\mathrm{W}},
\end{equation}
meanwhile leaving all other definitions and discussions unchanged.
However, for both PMA and IPMA cases, it is hopelessly complicated in mathematics if finite $a$ and $b$ coexist.
In the following subsections, we provide explicit analytics on finite enlargement of Walker limit
in the presence of either finite $a$ (FL-SOT) or finite $b$ (i-DMI).

\subsection{Coexistence of STT and FL-SOT}\label{WalkerBreakdownSuppression:OnlyFLSOT}
\begin{figure}[htbp]
	\centering
	\includegraphics[width=0.72\textwidth]{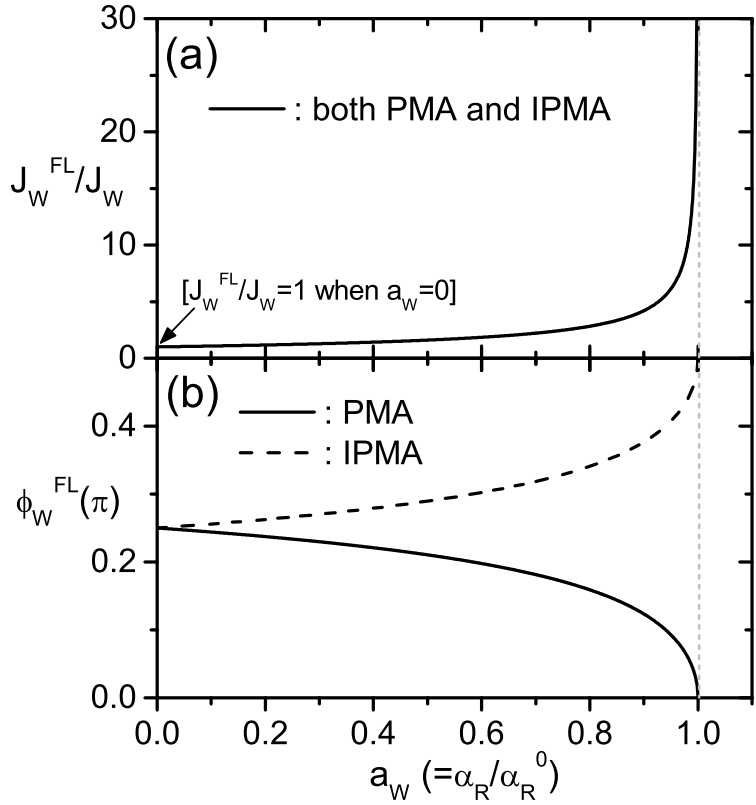}
	\caption{Walker breakdown suppression by FL-SOT as the function of parameter 
		$a_{\mathrm{W}}$ ($=\alpha_{\mathrm{R}}/\alpha_{\mathrm{R}}^0$ when Rashba SOT solely contributes to FL-SOT)
		in the absence of i-DMI:
		(a) Enlarged Walker limits normalized by $J_{\mathrm{W}}$ which are the same for PMA and 
		IPMA cases under the same $a_{\mathrm{W}}$. 
		(b) The azimuthal angle at which the enlarged Walker limit achieves. The solid and dashed
		curves are for PMA and IPMA systems, respectively. The gray short-dashed line indicates the occurrence of UAWB
	    when $a_{\mathrm{W}} \ge 1$. In all calculations we have set $\eta\mathrm{sgn}(\alpha-\beta)=-1$.}\label{fig2}
\end{figure}

First we set $b=0$ which means the i-DMI is absent.
For PMA systems, we set $\cos\phi\equiv u\in[-1,+1]$ and define a new function
$\mathcal{F}(u)$ from \eref{Je_JW_only_FL_SOTs} as
\begin{equation}\label{F_u_only_FL_SOTs}
\mathcal{F}(u)\equiv\left(j_a/J_{\mathrm{W}}\right)^2 =4u^2(1-u^2)/(1-a_{\mathrm{W}}u)^2.
\end{equation}
Obviously, $\mathcal{F}(u)$ is nonnegative. It equals to zero when and only when $u=0$ or $\pm1$.
To find its extrema, by setting $\mathcal{F}'(u)=0$
one obtains four extremal sites:
\begin{equation}\label{dF_du_4roots_only_FL_SOTs}
u_{k=1,2,3}=\frac{1}{3a_{\mathrm{W}}}\left[2+4\cos\frac{\theta+(2k-3)\pi}{3}\right],\;\; \theta=\arccos\left(\frac{27}{16}a_{\mathrm{W}}^2-1\right),
\end{equation}
and $u_4=0$.
Obviously $u_4$ is the minimum point which can be verified by $\mathcal{F''}(u_4)>0$.
When $a_{\mathrm{W}}\rightarrow 0^+$, direct calculation yields that
$u_1\approx 2/a_{\mathrm{W}}-a_{\mathrm{W}}/4\rightarrow +\infty$ and
$u_{2,3}\approx \pm 1/\sqrt{2}+a_{\mathrm{W}}/8\rightarrow \pm 1/\sqrt{2}$,
implying that $u_{2,3}$ are the two maximum points satisfying $-1<u_3<0<u_2<1$.
Also, it's easy to check $\mathcal{F}(u_2)>\mathcal{F}(u_3)>1$,
thus one has
\begin{equation}\label{WB_only_FL_SOTs}
J_{\mathrm{W}}^{\mathrm{FL}}=J_{\mathrm{W}}\sqrt{\mathcal{F}(u_2)},\quad \phi=\arccos u_2,
\end{equation}
which explicitly gives the enlarged Walker limit by the FL-SOT.
To be more intuitive, we plot $J_{\mathrm{W}}^{\mathrm{FL}}$ and
the corresponding $\phi(=\arccos u_2)$ in \fref{fig2}a and 2b, which perfectly
reproduce existing numerics (for example figure 3a in Ref. \cite{Seo_JMMM_2012}).

Next we calculate the divergent behavior of $J_{\mathrm{W}}^{\mathrm{FL}}$
when $a_{\mathrm{W}}\rightarrow 1^-$.
Define $\theta_0\equiv\arccos(27\cdot 1^2/16-1)=\arccos(11/16)$, which
satisfies $\cos\left(\theta_0/3+\pi/3\right)=1/4$.
Suppose $1-a_{\mathrm{W}}\equiv \epsilon\rightarrow 0^+$,
it turns out $u_2=1-\epsilon+o(\epsilon)$ from series expansions.
Putting $u_2$ back into equation \eref{WB_only_FL_SOTs}, after standard series-expansion calculation one gets
\begin{equation}\label{WB_aW_1_only_FL_SOTs}
J_{\mathrm{W}}^{\mathrm{FL}}\approx\sqrt{2}J_{\mathrm{W}}\cdot\left(1-a_{\mathrm{W}}\right)^{-1/2},
\end{equation}
with the corresponding $\phi$ satisfying
\begin{equation}\label{phi_aW_1_only_FL_SOTs}
 \phi\approx\sqrt{2}\cdot\left(1-a_{\mathrm{W}}\right)^{1/2}.
\end{equation}
Equations \eref{WB_aW_1_only_FL_SOTs} and \eref{phi_aW_1_only_FL_SOTs}
perfectly describe the asymptotic behaviors in \fref{fig2}a as $a_{\mathrm{W}}\rightarrow 1^-$.
This confirms the UAWB under appropriate combination of 
system parameters ($H_{\mathrm{K}}$ and $H_{\mathrm{FL}}$).
Also, when $a_{\mathrm{W}}\rightarrow 1^-$ one has
$u_3\rightarrow (1-\sqrt{5})/2\approx-0.618$ and
$\sqrt{\mathcal{F}(u_3)}\rightarrow 2^{-3/2}(\sqrt{5}-1)^{5/2}\approx 0.6006$.
This is a local maximum and will be abandoned as $|j_a|\rightarrow J_{\mathrm{W}}^{\mathrm{FL}}$.

Furthermore, in equation \eref{Je_JW_only_FL_SOTs} for fixed $\alpha$,
$\beta$ [thus fixed $\mathrm{sgn}(\alpha-\beta)$] and $j_a$,
the ``$\eta\rightarrow -\eta$" transformation is equivalent to
``$\phi\rightarrow\pi-\phi$". 
This means the sign-inversion in topological charge
does not change the polarity of a stable traveling wave.

To make the above mathematics more physical, here we take a limit case in which
the Rashba effective field\cite{Zhang_PRB_2008,Zhang_PRB_2009}
\begin{equation}\label{Effective_Rashba_field}
\mathbf{H}_{\mathrm{R}}=\lambda\left(\mathbf{e}_z \times \hat{\mathbf{J}}\right),\quad \lambda\equiv \frac{\alpha_{\mathrm{R}}P}{\mu_{\mathrm{B}} M_s}j_a,
\end{equation}
solely contributes to $H_{\mathrm{FL}}$.
Here $\alpha_{\mathrm{R}}$ is the Rashba parameter describing the Rashba SOC strength.
The resulting $\lambda$ is the conversion factor from current density to the Rashba field and
is about $10^{-8}\sim 10^{-9}\ \mathrm{T\ cm^2/A}$\cite{Miron_Nmat_2010,Seo_APL_2010}.
Then the definition of $a_{\mathrm{W}}$ in equation \eref{a_W_only_FL_SOTs} can 
be rewritten as
\begin{equation}\label{a_W_redefinition}
a_{\mathrm{W}}=\frac{\alpha_{\mathrm{R}}}{\alpha_{\mathrm{R}}^0},\quad \alpha_{\mathrm{R}}^0\equiv\frac{|\alpha-\beta|}{\alpha}\cdot\frac{1}{\Delta}\cdot \frac{2\mu_{\mathrm{B}}^2}{\pi e \gamma}.
\end{equation}
For magnetic parameters of Co-Ni FMM which is a typical 
PMA material (see the first column of \Tref{table1}, adopted from Ref. \cite{Risinggad_PRB_2017}),
one has $\alpha_{\mathrm{R}}^0=9.54$ $\mathrm{meV\cdot nm}$. 
Hence the above result means that for strong enough Rashba SOC ($|\alpha_{\mathrm{R}}|\ge \alpha_{\mathrm{R}}^0$),
the UAWB occurs, while for $|\alpha_{\mathrm{R}}|<\alpha_{\mathrm{R}}^0$
the Walker limit is finitely enlarged as described in equation \eref{WB_only_FL_SOTs}.
For this reason, the horizontal axis of \fref{fig2} can also be set as ``$\alpha_{\mathrm{R}}/\alpha_{\mathrm{R}}^0$".

In addtion, the critical condition ``$a_{\mathrm{W}}=1$" leads to 
``$\alpha_{\mathrm{R}}^{\mathrm{th}}\propto |\beta/\alpha-1|$" relationship
in the threshold above which no Walker breakdown occurs, thus explicates existing numerical
simulations, for example figure 3b in Ref. \cite{Seo_JMMM_2012}.

At last, for IPMA systems all the definitions and discussion are the same except for that 
the enlarged Walker limit achieves at $\phi=\arcsin u_2$.
When $a_{\mathrm{W}}\rightarrow 1^-$ the asymptotic behavior of $\phi$ turns to
\begin{equation}\label{phi_aW_1_only_FL_SOTs_IPMA}
\phi\approx\pi/2-\sqrt{2}\cdot\left(1-a_{\mathrm{W}}\right)^{1/2}.
\end{equation}
On the other hand, for magnetic parameters of permalloy FMM which is a typical 
IPMA material (see the second column of \Tref{table1}, adopted from Ref. \cite{Seo_APL_2012}),
one has a smaller critical Rashaba parameter $\alpha_{\mathrm{R}}^0=0.64$ $\mathrm{meV\cdot nm}$ 
due to the relatively large wall width. This means that in IPMA systems,
the UAWB is more likely to occur. 

\begin{table}
	\caption{\label{table1}Magnetic parameters used for numerical calculation and estimations. The first (second) column 
		comes from Co-Ni (Py) which constitutes a typical PMA (IPMA) FMM layer in trilayers.}
	\footnotesize
	\begin{tabular}{@{}llll}
		\br
		Magnetic parameter & Co-Ni (from Ref. \cite{Risinggad_PRB_2017}) & Py (from Ref. \cite{Seo_APL_2012}) & unit  \\
		\mr
		gyromagnetic rato ($\gamma$): & 1.76 & 1.76 & $10^{11}$ Hz/T \\
		saturation magnetization ($M_s$): & 1.0 & 0.8 & $10^6$ A/m \\
		FMM length : &   & 2000 & nm\\
		FMM width : &   & 80 & nm\\
		FMM thickness ($t_{\mathrm{F}}$): & 1.2 & 4 & nm\\
		hard axis anisotropy ($k_{\mathrm{H}}$): & $1/\pi$ & 0.8555 (=$D_y-D_x$) &   \\
		domain wall width ($\Delta$): & 4 & 30 & nm\\
		i-DMI magnitude ($D$): & $-1.4$ &   &  mJ/$\mathrm{m^2}$ \\
		Gilbert damping ($\alpha$): & 0.25 & 0.02 &     \\
		spin polarization ($P$): & 0.5 & 0.7 &  \\
		nonadiabacity parameter ($\beta$): & 0.5 & 0.01 &  \\
		Rashba parameter ($\alpha_{\mathrm{R}}$): & 6.3 &  &  $\mathrm{meV\cdot nm}$ \\
		spin Hall angle ($\theta_{\mathrm{SH}}$): & 0.1 & 0.1 &  \\
		
		\br
	\end{tabular}\\
	
\end{table}
\normalsize

\subsection{Coexistence of STT and i-DMI}\label{WalkerBreakdownSuppression:OnlyDMI}
In this subsection, we study whether i-DMI itself can lead to 
UAWB, thus we have $a=0$ and $b\ne 0$.
As usual, we first focus on PMA systems.
By setting $\cos\phi\equiv s\in[-1,+1]$,
another function $\mathcal{G}(s)$ can be defined from \eref{phi_dot_0_only_FL_SOTs} as
\begin{equation}\label{G_s_only_FL_SOTs}
\mathcal{G}(s)\equiv\left(j_a/J_{\mathrm{W}}\right)^2 = (1-s^2)(2s-b)^2.
\end{equation}
$\mathcal{G}(s)$ is also nonnegative and only equals to zero when
$s=\pm1$ or $b/2$.
By requiring $\mathcal{G}'(s)=0$, three extremal sites are obtained
\begin{equation}\label{dG_ds_3roots_only_FL_SOTs}
s_1=\frac{b}{2},\; s_2=\frac{b+\sqrt{b^2+32}}{8},\; s_3=-\frac{4}{b+\sqrt{b^2+32}}.
\end{equation}

\begin{figure}[htbp]
	\centering
	\includegraphics[width=0.72\textwidth]{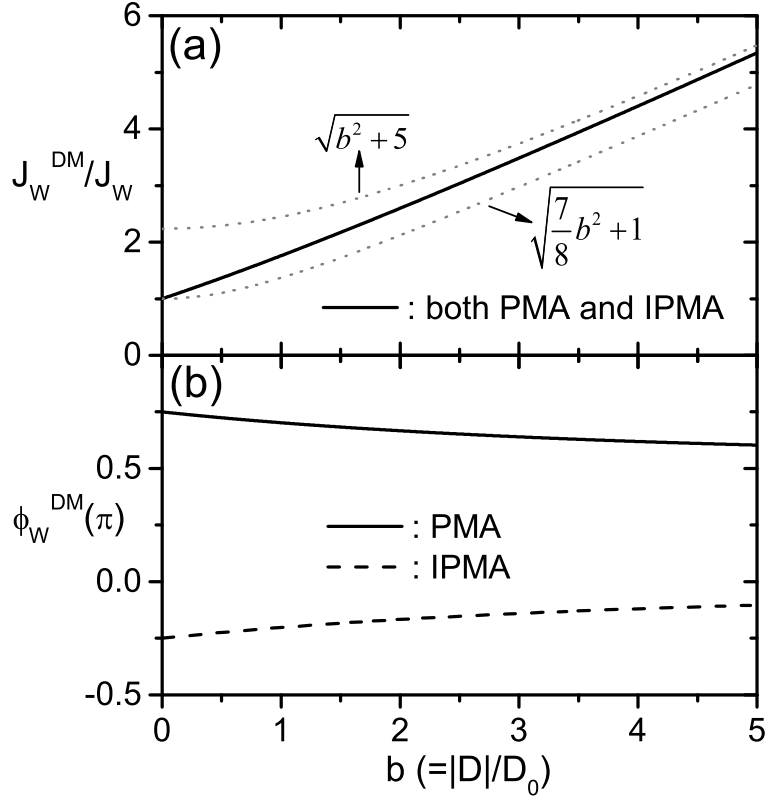}
	\caption{Walker breakdown suppression by pure i-DMI as the function of parameter 
		$b$ ($=|D|/D_0$) in the absence of any SOTs:
		(a) Enlarged Walker limits normalized by $J_{\mathrm{W}}$ which are the same for PMA and 
		IPMA systems under the same $b$. The gray dotted curves indicate the upper and lower boundaries.
		(b) The azimuthal angle at which the enlarged Walker limit achieves. The solid and dashed
		curves are for PMA and IPMA systems, respectively. 
		 In all calculations, $\eta\mathrm{sgn}(\alpha-\beta)=-1$ and $\eta\mathrm{sgn}(D)=+1$.}\label{fig3}
\end{figure}

Standard calculus tells us that function $\mathcal{G}(s)$ always
approaches maximum at $s=s_3$ when $b>0$ ($s=s_2$ for $b<0$).
Then the modified Walker limit [maximum absolute value of the right hand side
of equation \eref{G_s_only_FL_SOTs}] and the corresponding $\phi$ reads
\begin{equation}\label{WB2_only_FL_SOTs}
J_{\mathrm{W}}^{\mathrm{DM}}=J_{\mathrm{W}}\sqrt{\mathcal{G}(s_3)}, \quad \phi=\arccos s_3.
\end{equation} 
Meanwhile, simple calculation yields that $7b^2/8+1<\mathcal{G}(s_3)<b^2+5$, which
confirms the Walker breakdown suppression effect at finite $b$ (i.e. i-DMI).
However since no singularities appear, UAWB 
does not occur under the pure action of finite i-DMI.

For IPMA systems, again the definitions and discussions are similar. 
The enlarged Walker limit can be described by equation \eref{WB2_only_FL_SOTs}
except for that the corresponding extremum point locates at $\phi=\arcsin s_3$.

Also, one should note that although the enlarged Walker limits in
PMA and IPMA cases share the same analytical form, they will take
different value under the same i-DMI strength $D$.
To see this, we rewrite the definition of $b$ in equation \eref{a_b_definition} as
\begin{equation}\label{b_definition_rewrite}
b=\frac{\eta D}{D_0},\quad D_0\equiv \frac{\mu_{0}M_s H_{\mathrm{K}}\Delta}{\pi}.
\end{equation}
For Co-Ni FMM (PMA), one has $D_0=0.51 \mathrm{mJ/m^2}$. 
While for Py FMM (IPMA), one gets $D_0=6.57 \mathrm{mJ/m^2}$ due to the larger wall width therein.
Thus generally i-DMI induces stronger Walker breakdown suppression in PMA systems
under the same i-DMI strength $D$.
In \fref{fig3}a and 3b, we plot $J_{\mathrm{W}}^{\mathrm{DM}}$ and the corresponding
$\phi$ for $0\le b\le 5$ when $\eta\mathrm{sgn}(D)=+1$.
Clearly, the behaviors of enlarged Walker limits in \fref{fig2}a and \fref{fig3}a are totally different.

The qualitative role of FL-SOTs and i-DMI has been extensively discussed and now is clear. 
The driving current pulls the magnetization out of the 
easy plane, while the demagnetization field ($H_{\mathrm{K}}$) tends to prevent this from happening,
leading to the classical Walker limit $H_{\mathrm{W}}$. The extra effective field from SOC ($H_{\mathrm{FL}}$)
and i-DMI in $\mathbf{e}_y$ axis (hold for both PMA and IPMA cases) also helps to prevent 
the magnetization from leaving the easy plane, thus extending
the traveling-wave region of walls. This is the physical origin of Walker breakdown suppression.
Our analytics here provides detailed and solid foundation for the above physical picture.

\section{Mobility change by ADL-SOT}\label{MobilityChangeByADLSOT}
To understand the mobility change and even its sign-inversion of 
N\'{e}el walls in FMM layer of trilayers in quite a lot experiments and simulations,
the ADL-SOT must be appended. 
With the coexistence of STT, FL-SOT, ADL-SOT and i-DMI, the traveling-wave ansatz
in equation \eref{Walker_profile_traveling_wave} is again selected as the start-point of investigation.
We will show that unlike the similarity in ``Walker breakdown suppression" part, 
the mobility changes by ADL-SOT take quite different forms for PMA and IPMA cases.

\subsection{General framework under the coexistence of STT, FL-SOT, ADL-SOT and i-DMI}\label{MobilityChangeByADLSOT:GeneralFramework}
As usual, we first concentrate on PMA cases.
Under the traveling-wave ansatz, one has
\begin{equation}\label{AB_tilde_FL_and_ADL_SOTs}
\eqalign{
	\tilde{A}=\frac{2\eta D}{\mu_{0} M_s\Delta}\sin^2\theta\sin\phi-\frac{H_{\mathrm{K}}}{2}\sin\theta\sin 2\phi +H_{\mathrm{FL}}\cos\phi \cr
	\qquad \qquad +H_{\mathrm{ADL}}\cos\theta\sin\phi,\cr
	\tilde{B}=H_{\mathrm{ADL}}\cos\phi-H_{\mathrm{FL}}\cos\theta\sin\phi.
}
\end{equation}
After putting back into the scalar LLG equations and integrating over the whole strip, it turns out
\begin{equation}\label{v_phi_dot_FL_and_ADL_SOTs}
	\eqalign{
		\frac{v(t)}{\Delta(\phi)}=\frac{\eta\gamma \left(\mathcal{H}+\alpha\pi H_{\mathrm{ADL}}\cos\phi\right)}{2(1+\alpha^2)} - \frac{1+\alpha\beta}{1+\alpha^2}\frac{B_J}{\Delta(\phi)}, \cr
		\dot{\phi}=-\frac{\alpha\gamma \left(\mathcal{H} -\pi H_{\mathrm{ADL}}\cos\phi/\alpha\right)}{2(1+\alpha^2)} +\frac{\alpha-\beta}{1+\alpha^2}\frac{\eta B_J}{\Delta(\phi)},
	}
\end{equation}
with the same functional $\mathcal{H}(H_{\mathrm{K}},H_{\mathrm{FL}},D,\phi)$ 
defined in equation \eref{H_functional_only_FL_SOTs}.
Note that the structure of equation \eref{v_phi_dot_FL_and_ADL_SOTs} is different from that of
equation \eref{v_phi_dot_Walker_traveling_wave} due to the presence of $H_{\mathrm{ADL}}-$terms.

By requiring $\dot{\phi}=0$ in equation \eref{v_phi_dot_FL_and_ADL_SOTs}, one gets
\begin{equation}\label{phi_dot_0_FL_and_ADL_SOTs}
\sin 2\phi-b\sin\phi-c\cos\phi=\eta \mathrm{sgn}(\alpha-\beta)\frac{j_a}{J_\mathrm{W}}, \;\; c\equiv
\pi\frac{H_{\mathrm{FL}}}{H_{\mathrm{K}}}\left(1+\frac{H_{\mathrm{ADL}}}{\alpha H_{\mathrm{FL}}}\right),
\end{equation}
Since $H_{\mathrm{FL}}$ and $H_{\mathrm{ADL}}$ are both proportional to $j_a$, 
we set $\left(H_{\mathrm{FL}}\right)_{\mathrm{W}}$ and
$\left(H_{\mathrm{ADL}}\right)_{\mathrm{W}}$ as the absolute effective-field strengths of FL- and ADL-SOTs
when $|j_a|=J_{\mathrm{W}}$, respectively.
Then after defining
\begin{equation}\label{b_W_FL_and_ADL_SOTs}
c_{\mathrm{W}}\equiv \pi\left[\left(H_{\mathrm{FL}}\right)_{\mathrm{W}}+\alpha^{-1}\left(H_{\mathrm{ADL}}\right)_{\mathrm{W}}\right]/ H_{\mathrm{K}}>0,
\end{equation}
equation \eref{phi_dot_0_FL_and_ADL_SOTs} is rewritten as
\begin{equation}\label{Je_JW_FL_and_ADL_SOTs}
\frac{j_a}{J_{\mathrm{W}}} =\frac{\sin 2\phi-b\sin\phi}{\eta \mathrm{sgn}(\alpha-\beta)+c_{\mathrm{W}}\cos\phi}.
\end{equation}
The rest discussion on Walker breakdown suppression is the same as 
those in \sref{WalkerBreakdownSuppression:GeneralFramework} to \sref{WalkerBreakdownSuppression:OnlyDMI}.
We define the enlarged Walker limit as $J_{\mathrm{W}}^{\mathrm{all}}$.
For $|j_a|\le J_{\mathrm{W}}^{\mathrm{all}}$, from equation \eref{v_phi_dot_FL_and_ADL_SOTs}, 
the wall velocity reads
\begin{equation}\label{v_all3_PMA}
v=-\frac{\beta}{\alpha}B_J+\frac{\eta\gamma\pi\Delta}{2\alpha}H_{\mathrm{ADL}}\cos\phi.
\end{equation}

As for IPMA systems, after similar discussion, the existence condition of traveling-wave mode
, $\dot{\phi}=0$, provides
\begin{equation}\label{phi_dot_0_all3_IPMA}
\sin 2\phi-b\cos\phi+c\sin\phi=\eta \mathrm{sgn}(\alpha-\beta)\frac{j_a}{J_\mathrm{W}},
\end{equation}
and the corresponding wall velocity reads
\begin{equation}\label{v_all3_IPMA}
v=-\frac{\beta}{\alpha}B_J-\frac{\eta\gamma\pi\Delta}{2\alpha}H_{\mathrm{ADL}}\sin\phi.
\end{equation}

\subsection{Mobility change in PMA systems}\label{MobilityChangeByADLSOT:PMA}
In principle, to obtain the wall velocity in traveling-wave mode, one should solve $\phi$
from its existence condition [for PMA sytems, equation \eref{phi_dot_0_FL_and_ADL_SOTs}]
and then put it into the velocity formula [see equation \eref{v_all3_PMA}].
However the calculation process is hopelessly complicated.

Inspired by the asymptotic approach\cite{AGoussev_PRB,AGoussev_Royal, jlu_prb_2016,jlu_SciRep_2017,jlu_Nanomaterials_2017},
we consider the case where $|j_a|\ll J_{\mathrm{W}}^{\mathrm{all}}$ thus
the wall must be in traveling-wave mode ($\dot{\phi}=0$) and
$\phi$ is not far from its static position ($\phi_0=\arccos[-\eta\mathrm{sgn}(D)]$).
Next we introduce the azimuthal deviation (see \fref{fig1}a)
\begin{equation}\label{psi_definition_all3_PMA}
\psi\equiv\phi-\phi_0.
\end{equation}
Since $|\psi|\ll 1$, thus $\sin\phi\approx[-\eta\mathrm{sgn}(D)]\psi$,
$\cos\phi\approx -\eta\mathrm{sgn}(D)$ and $\sin 2\phi\approx 2\psi$.
Putting them into equation \eref{phi_dot_0_FL_and_ADL_SOTs}, the azimuthal deviation $\psi$ can be
solved as
\begin{equation}\label{psi_solved_all3_PMA}
\psi\approx \eta\frac{\mathrm{sgn}(\alpha-\beta)-\mathrm{sgn}(D)c_{\mathrm{W}}}{2+|D|/D_0}\frac{j_a}{j_{\mathrm{W}}}.
\end{equation}
Meanwhile the wall velocity in equation \eref{v_all3_PMA} becomes
\begin{equation}\label{v_all3_PMA_small_ja}
v=-\frac{\beta}{\alpha}\frac{\mu_{\mathrm{B}} P}{e M_s}j_a+\frac{-\mathrm{sgn}(D)\gamma\pi\Delta}{2\alpha}H_{\mathrm{ADL}}.
\end{equation}
Since $H_{\mathrm{ADL}}$ is proportional to $j_a$, thus the wall mobility can be changed.

One extreme case is that ADL-SOT is induced solely by SHE.
Thus $H_{\mathrm{ADL}}=H_{\mathrm{SHE}}=\hbar\theta_{\mathrm{SH}}/(2\mu_{0}e M_s t_{\mathrm{F}})$.
Then equation \eref{v_all3_PMA_small_ja} turns to
\begin{equation}\label{v_all3_PMA_small_ja_final}
\frac{v_{\mathrm{PMA}}}{v_{\mathrm{STT}}}=1+\frac{\mathrm{sgn}(D)\theta_{\mathrm{SH}}}{\theta_{\mathrm{SH}}^0}, \quad v_{\mathrm{STT}}\equiv -\frac{\beta}{\alpha} B_J,
\end{equation}
in which $\theta_{\mathrm{SH}}^0\equiv 2\beta P \mu_{0} e t_{\mathrm{F}}/(\gamma\pi m_e \Delta)$
and $m_e$ is the electron mass. 
As long as $[-\mathrm{sgn}(D)]\theta_{\mathrm{SH}}>\theta_{\mathrm{SH}}^0$, the motion direction
of the wall will be reversed ($v_{\mathrm{PMA}}/v_{\mathrm{STT}}<0$).
Note that to achieve this, not only the strength but also the sign of spin Hall angle should be specified.
This former is controlled by the ADL-SOT strength while the latter is selected by the sign of i-DMI parameter.
For Co-Ni FMM, one has $\mathrm{sgn}(D)=-1$ and $\theta_{\mathrm{SH}}^0\approx 0.048<0.1=[-\mathrm{sgn}(D)]\theta_{\mathrm{SH}}$. 
Therefore N\'{e}el walls in this trilayer will move along charge current direction.
In addition, by carefully arranging magnetic parameters [$\beta$, $\Delta$, $t_{\mathrm{F}}$ and $\mathrm{sgn}(D)$]
of trilayer systems, $|v_{\mathrm{PMA}}/v_{\mathrm{STT}}|$ can be much higher than 1.
This will help to explain the relatively high velocity of domain walls 
in $\mathrm{Pt(3\;nm)/Co(0.6\;nm)/AlO_x(2\;nm)}$ trilayer ($\sim 400\ \mathrm{m/s}$ when $j_a\sim 10^{8}\;\mathrm{A/cm^2}$)\cite{Miron_Nmat_2011}.

\subsection{Mobility change in IPMA systems}\label{MobilityChangeByADLSOT:IPMA}
As indicated in \sref{Modeling:StaticWall}, in IPMA cases the i-DMI does not select wall polarity
thus $\phi_0=k\pi,k\in\mathbb{Z}$, i.e. $\cos\phi_0=(-1)^k$.
Again we introduce the azimuthal deviation $\psi\equiv \phi-\phi_0$ under small current density.
Since $|\psi|\ll 1$, thus $\sin\phi\approx (-1)^k\psi$, $\cos\phi\approx (-1)^k$ and $\sin 2\phi\approx 2\psi$.
Putting them into equation \eref{phi_dot_0_all3_IPMA}, one has
\begin{equation}\label{psi_solved_FL_and_ADL_SOTs}
\psi\approx \eta\left[\frac{(\alpha-\beta) B_J}{\gamma \Delta H_{\mathrm{W}}}+\frac{(-1)^k D}{D_0}\right]\cdot\left[2+\frac{(-1)^k \pi}{H_{\mathrm{K}}}\left(H_{\mathrm{FL}}+\frac{H_{\mathrm{ADL}}}{\alpha}\right)\right]^{-1}.
\end{equation}
Meanwhile the wall velocity is reduced to
\begin{equation}\label{v_all3_IPMA_small_ja}
v=-\frac{\beta}{\alpha}B_J-\frac{\eta\gamma\pi\Delta}{2\alpha}H_{\mathrm{ADL}}(-1)^k\psi,
\end{equation}
which is complicated due to the coexistence of $H_{\mathrm{FL}}$, $H_{\mathrm{ADL}}$ and $D$.

Next we consider an extreme case where the i-DMI is neglected.
After simple algebra, the wall velocity in equation \eref{v_all3_IPMA_small_ja} becomes
\begin{equation}\label{vt_FL_and_ADL_SOTs}
\frac{v_{\mathrm{IPMA}}}{v_{\mathrm{STT}}}=\frac{1+\frac{(-1)^k\pi H_{\mathrm{ADL}}}{2 H_{\mathrm{K}}}\left(\zeta+\beta^{-1}\right)}{1+\frac{(-1)^k\pi H_{\mathrm{ADL}}}{2 H_{\mathrm{K}}}\left(\zeta+\alpha^{-1}\right)},
\end{equation}
with $\zeta\equiv H_{\mathrm{FL}}/H_{\mathrm{ADL}}$ and can be assumed positive without losing generality.
{\it Obviously, the presence of $H_{\mathrm{ADL}}$ as well as the condition $\alpha\ne\beta$
provides us the possibility of changing wall mobility.}
Interestingly, the direction of wall motion can even be reversed ($v_{\mathrm{IPMA}}/v_{\mathrm{STT}}<0$)
under the following condition
\begin{equation}\label{v_reversal_1_FL_and_ADL_SOTs}
\frac{1}{\zeta+\max(\frac{1}{\alpha},\frac{1}{\beta})}<\frac{(-1)^{k+1}\pi H_{\mathrm{ADL}}}{2 H_{\mathrm{K}}}<\frac{1}{\zeta+\min(\frac{1}{\alpha},\frac{1}{\beta})}.
\end{equation}
In addition, the wall halts at $(-1)^{k+1}\pi H_{\mathrm{ADL}}=2 H_{\mathrm{K}}/(\zeta+\beta^{-1})$
and a velocity divergence occurs at $(-1)^{k+1}\pi H_{\mathrm{ADL}}=2 H_{\mathrm{K}}/(\zeta+\alpha^{-1})$.

\Eref{v_reversal_1_FL_and_ADL_SOTs} shows that only walls with polarity satisfying $(-1)^{k+1}H_{\mathrm{ADL}}>0$
can be reversed from electron flow to charge current direction, which well
explains the ``polarity sensitivity" phenomena in IPMA systems.
On the other hand, in real IPMA materials $\alpha,\beta\ll 1$.
Therefore the current density under which equation \eref{v_reversal_1_FL_and_ADL_SOTs} holds
can be small enough to ensure the approximation for obtaining equation \eref{psi_solved_FL_and_ADL_SOTs},
thus makes the whole deduction coherent.

\begin{figure}[htbp]
	\centering
	\includegraphics[width=0.7\textwidth]{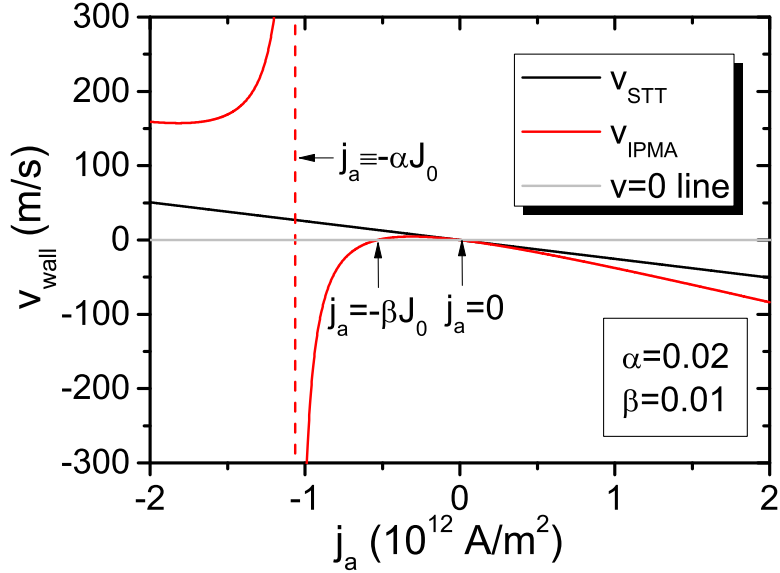}
	\caption{(Color online) Mobility change and motion-direction reversal in the presence of
		ADL-SOT for $\alpha>\beta$. Here $\alpha=0.02$, $\beta=0.01$ and $D=0$. Other
		magnetic parameters are taken as those in the second column of \Tref{table1}.
		The solid black line and red curve indicate $v_{\mathrm{STT}}$ and $v_{\mathrm{IPMA}}$,
		respectively. The vertical dash line is the $j_a\equiv -\alpha J_0$ line where
		divergence in $v_{\mathrm{IPMA}}$ occurs.}\label{fig4}
\end{figure}

To numerically check our analytics, in the simplest case we set
$H_{\mathrm{FL}}=0$ (thus $\zeta=0$) and suppose that $H_{\mathrm{ADL}}$ 
solely comes from SHE, which is exactly the case in Ref. \cite{Seo_APL_2012}.
Under the magnetic parameters in the second column of \Tref{table1},
the ``$v_{\mathrm{IPMA}}(v_{\mathrm{STT}})\sim j_a$" curves are plotted in \fref{fig4}.
The black line indicates the linear dependence of $v_{\mathrm{STT}}$ on $j_a$,
while the red curve represent the wall velocity $v_{\mathrm{IPMA}}$ when only SHE-induced ADL-SOT is considered.
The reversal region is $-\alpha J_0<j_a<-\beta J_0$ with $\alpha=0.02$, $\beta=0.01$
and $J_0\equiv 4\mu_{0} e k_{\mathrm{H}} M_s^2 t_{\mathrm{F}}/(\pi\hbar\theta_{\mathrm{SH}})=5.32\times 10^{13}$ $\mathrm{A/m^2}$.
The wall-halt current is $-\beta J_0=-5.32\times 10^{11}$ $\mathrm{A/m^2}$
and the velocity-divergence current is $-\alpha J_0=-1.064\times 10^{12}$ $\mathrm{A/m^2}$.
All these results reproduce very well the ``$\theta_{\mathrm{SH}}=+0.1$" case in figure 3a of Ref. \cite{Seo_APL_2012}.
In addition, the positively divergent part for ``$j_a<-\alpha J_0$" in \fref{fig4}
indicates the possibility of ``velocity boosting" in the original direction (electron flow) 
by SHE-induced ADL-SOTs.
There are few reports in literatures about this and
should be worth of more efforts in both simulations and experiments.
In particular, it may help to explain the unexpected high
velocities of domain walls in permalloy layer of multilayer films in some early works\cite{Hayashi_PRL_2007}.

\subsection{Discussions}
All analytics in this work are performed based on the
traveling-wave ansatz in equation \eref{Walker_profile_traveling_wave}.
One must bear in mind that it is rigorous only in the
absence of any SOTs and i-DMI, otherwise in principle it fails to provide the rigorous
wall profile. However, it may serve as a ``not-bad" approximation of the actual
magnetization texture in trilayers.
This has been numerically tested in Ref. \cite{Risinggad_PRB_2017}.

Second, as this ansatz can not hold everywhere
along the long axis of strip, to obtain the collective behaviors we then integrate
it over $r\in(-\infty,+\infty)$ which is transferred to the integration
of $\theta\in(0,\pi)$. However, when $j_a$ increases, effective
transverse fields from SOTs and i-DMI will pull the magnetization in two
faraway domains away from strip axis. Then the integration of
$r\in(-\infty,+\infty)$ should be converted to
that of $\theta$ over $(\theta_0,\pi-\theta_0)$, where
$\theta_0$ is positively correlated with $j_a$ with some complicated mathematical dependence.
For simplicity, we have not considered this $\theta_0$ in the above sections.
Further investigation on this issue is out of the scope of this work.

At last, by ``small quantity analysis" we succeed in explaining
the mobility change in both PMA and IPMA systems. 
In particular, the sign-inversion of mobility
as well as the ``polarity sensitivity" therein is recovered analytically.
In real experiments, the motion-direction-reversal behavior is observed
in a relatively wide range of $j_a$ (thus $H_{\mathrm{ADL}}$).
This should not be regarded as a contradiction with equation \eref{v_reversal_1_FL_and_ADL_SOTs}
since it is obtained under the assumption $|j_a|\ll J_{\mathrm{W}}^{\mathrm{all}}$.
In fact, the necessity of ADL-SOTs for mobility sign-inversion,
as well as the polarity selection rule therein, should be the main focus in this subsection.
Also, this is one of the main reasons why this part of SOT is named as ``anti-damping-like"
since they can input energy into the system, not just dissipate it.

\section{Summary}
In this work, we analytically investigate the current-induced domain wall dynamics
in HM/FMM/Oxide trilayers with strong SOCs and i-DMI.
We show that in both PMA and IPMA systems,
FL-SOT can induce UAWB but i-DMI can not.
For moderate FL-SOT and arbitrary i-DMI strength, 
we provide analytical expressions of the finitely enlarged Walker limits.
On the other hand, the wall mobility change can be explained only when
ADL-SOT is included under the coexistence of STT, SOT and i-DMI.
In particular, for PMA systems strong enough spin Hall angle and appropriate sign of 
i-DMI parameter will lead to sign-inversion in wall mobility even under small enough current density,
while for IPMA systems this will only occur when current density falls into a finite range.
These analytical results provide insights not only for explaining existing experimental and numerical
data (in fact a numbers of them have been explained in the main text),
but also for the development of future domain-wall-based magnetic nanodevices.

\ack
This work is supported by the National Natural Science Foundation of
China (Grants Nos. 51671148, 11674251, 51601132 and 11374088).
J. Lu also acknowledges the support from the Natural Science Foundation
of Hebei Province, China (Grant No. A2014205080),
and the Science Foundation for The Excellent Youth Scholars of Educational Commission of Hebei Province,
China (Grant No. Y2012027).

\end{document}